\newif\ifhide
\documentclass[a4paper,fleqn]{cas-dc}

\usepackage[numbers]{natbib}

\usepackage{amsmath,amsfonts,amssymb,amsthm}
\usepackage{graphicx}
\usepackage{braket}
\usepackage{hyperref}
\usepackage{color}
\usepackage{soul}

\renewcommand{\vec}[1]{\boldsymbol{#1}}
\newcommand{\mtx}[1]{\boldsymbol{\mathsf{#1}}}

\newcommand{\bigket}[2]{\left|\!\!\begin{array}{#1}#2\end{array}\!\!\right\rangle}

\newif\ifshowfigures
\showfigurestrue % uncomment this line to reveal the figures

\ifhide

\else

\fi

\def\tsc#1{\csdef{#1}{\textsc{\lowercase{#1}}\xspace}}
\tsc{SRT}
\tsc{BRL}

\begin{document}

\let\WriteBookmarks\relax
\def\floatpagepagefraction{1}
\def\textpagefraction{.001}

\shorttitle{}
\shortauthors{S. Turner and B. La Cour}

\title[mode = title]{Pinching operators for approximating multiphoton entangled states}                      

\author[1]{Skylar R. {Turner}}[]
\ead{srturner@hrl.com}
\credit{Writing - Original draft preparation, Visualization, Formal analysis, Software}

\affiliation[1]{ organization={HRL Laboratories, LLC},
                     addressline={3011 Malibu Canyon Road}, 
                     city={Malibu},
                     state={CA},
                     postcode={90265}, 
                     country={USA}
}

\author[2]{Brian R. {La Cour}}[orcid=0000-0001-7899-0938]
\ead{blacour@arlut.utexas.edu}
\credit{Writing - review \& editing, Conceptualization, Project administration, Funding acquisition}
\cormark[1]

\affiliation[2]{ organization={Applied Research Laboratories, The University of Texas at Austin},
                     addressline={P.O. Box 9767}, 
                     city={Amsterdam},
                     postcode={78766-9767}, 
                     country={USA}
}
\cortext[cor1]{Corresponding author}

\begin{abstract}
We introduce the pinching operator, which extends the theory of squeezing operators to non-Gaussian operators, and use it to approximate $n$-photon entangled states using a pinched vacuum state and pinching tensor of rank $n$.  A simple recursion relation is derived for generating the Bogoliubov transformed creation and annihilation operators, which may be used to express the pinched state as a statistically equivalent set of nonlinearly transformed complex Gaussian random variables.  Using this representation, we compare low-order approximations of the pinched state to entangled multiphoton Fock states, such as Greenberger-Horne-Zeilinger (GHZ) and W states.  Using post-selection and a threshold detector model to represent non-Gaussian measurements, we find that this model is capable of producing states with a fidelity comparable to that of experimentally prepared multiphoton entangled states.  Our results show that it is possible to classically simulate large multiphoton entangled states to high fidelity within the constraints of finite detection efficiency.
\end{abstract}

\begin{keywords}
squeezing \sep entanglement \sep multiphoton \sep non-Gaussian states
\end{keywords}

\maketitle

%%%%%%%%%%%%%%%%%%%%%%%%%%%%%%%%%%%%%%%%%%%%%%%%%

\section{Introduction}

Squeezed states of light are an important feature of quantum optics that arises from the interactions of light with nonlinear optical media \cite{Walls1983,GhoshEtAl1986,Andersen2016}.  Squeezed coherent states had been considered theoretically for quite some time but were not realized experimentally until 1985 \cite{Kennard1927,Plebanski1965,Slusher1985}.  By contrast, two-photon squeezed states were first described theoretically in the 1970s and only later realized experimentally, in 1986 \cite{Klyshko1970,Yuen1976,Wu1986}.

Squeezed vacuum and coherent states are known to exhibit sub-Poissonian statistics \cite{Lee1990}, while multimode squeezed states arising from, say, parametric down conversion (PDC) can give rise to entanglement \cite{CavesSchumaker1985}.  Squeezed states of coherent light have numerous practical applications, such as high-precision measurements of gravitational waves and quantum key distribution \cite{Aasi2013,Derkach2020}.  Entangled squeezed states arising from parametric down conversion have numerous applications in quantum sensing \cite{Lawrie2019}, communication \cite{Slusher1990}, and computing \cite{Menicucci2006,Noh2020}.

Squeezed states are generally considered to be nonclassical due to their lack, in many cases, of a well defined Glauber-Sudarshan $P$ function that could be interpreted as a probability density function over coherent states \cite{Sperling2016}.  For example, Agarwal has shown that a bimodal vacuum state with nonzero squeezing does not admit such a $P$ function \cite{Agarwal1988}.  Similarly, Kim et al.\ have shown that for sufficiently squeezed thermal states the $P$ function takes negative values \cite{Kim1989}.  Notwithstanding, squeezed vacuum and coherent states, whether of single or multiple modes, may be described by a Gaussian, and hence positive, Wigner function that also serves as a probability density function for the quadratures of the state.  In particular, it is known that such states can be simulated efficiently by classical means \cite{Mari&Eisert2012,Bartlett2002}.  By contrast, all non-Gaussian pure states are known to have Wigner functions that exhibit some negativity \cite{Hudson1974,Soto1983,Walschaers2021}.

Experimental processes such as parametric downconversion and four-wave mixing can readily generate Gaussian biphoton entangled states, such as Bell pairs, to high fidelity.  However, the generation of non-Gaussian multiphoton entangled states (i.e., those with three or more photons) poses a greater challenge \cite{Erhard2020}.  Traditionally, interference and post-selection have been used to generate multiphoton entangled states from one or more biphoton entanglement sources \cite{Bouwmeester1999,Pan2001,Eibl2004}.  More recently, techniques such as cascaded sources and atomic vapors have been used to generate non-Gaussian multiphoton states directly \cite{Hubel2010,Wen2010}.  Such experiments are often difficult to perform and have generally low state generation rates.  There may therefore be some practical benefit to classical methods of simulating non-Gaussian multiphoton entangled states that are fast, scalable, and of comparable fidelity.

Prior work in the classical simulation of entangled states using multimode squeezed states has shown that the Gaussian nature of the quantum state, and hence of the Wigner function, can be represented faithfully by a complex Gaussian random vector with a nonzero pseudo covariance matrix, provided one is restricted to symmetrically ordered operators as observables.  This, of course, is well known.  More interestingly, if one uses a simple detector model based on amplitude threshold exceedances and applies post-selection on invalid outcomes, such as nondetection events, this representation can reproduce, perhaps surprisingly, many of the key qualitative features of a two-photon entangled state, such as violations of the Bell-CHSH inquality \cite{LaCour2021}.  It may seem surprising that such behavior can be exhibited by a classical simulation, but it is entirely possible within the constraints of nonideal detection efficiency \cite{Larsson2014}.  Motivated by these results, the present work seeks to extend this representation to arbitrary multiphoton entangled states.

Squeezed states may be defined by a squeezing operator $\hat{S} = \exp(\hat{A})$ acting on, say, the vacuum state, where $\hat{A} = -\hat{A}^\dagger$ is skew-Hermitian.  Prior work in the theoretical description of squeezed light has focused mainly on one or two spatial modes, for which $\hat{A}$ is a linear or quadratic function of the multimodal creation and annihilation operators.  Quadratic forms can be used to represent entangled photon pairs, as is the case for parametric down conversion, and have been routinely used to conduct Bell tests \cite{Kwiat1995}.  It must be understood, however, that the states produced in these experiments are not, in fact, Bell states but, rather, superpositions of the vacuum state (to zeroth order), a two-photon Bell state (to first order), and higher order Fock states, which Iskhakov et al.\ in have termed \emph{macroscopic Bell states} \cite{Iskhakov2011}.

Although squeezed states may consist of an arbitrary number of modes, their quadratic form limits them to describing no more than two-photon entangled states, at least to first order in the Fock decomposition.  Other, generalized notions of squeezed states have, however, been suggested.  Satyanarayana introduced the notion of \emph{generalized squeezed coherent states}, for which the single-mode squeezing operator is applied to a general Fock state, followed by an application of the displacement operator \cite{Satyanarayana1985}.  Nieto and Truax have described a generalization in which the single-mode squeezed states are eigenstates of linear combinations of creation and annihilation operators \cite{Nieto1993}.  At about the same time, Lo and Sollie introduced the notion of \emph{generalized multimode squeezed states}, for which squeezing may be performed on an arbitrary number of modes \cite{Lo1993}.  We note that for all such generalizations the form of $\hat{A}$ remains quadratic.  Hence, in the weak-squeezing regime such states can represent at most two entangled photons.

Higher-order nonlinearities can give rise to generalized squeezing with cubic and higher order forms for $\hat{A}$, although one must take care in defining them \cite{Gorska2014}.  When acted upon the vacuum state, the resulting quantum states are non-Gaussian and can be a resource for quantum computing.  Several such higher-order generalizations of squeezing have been introduced, all of them restricted a single mode.  For example, Braunstein and MacLachlan use the term \emph{generalized squeezing} to describe a special case of single-mode, multiphoton squeezing \cite{Braunstein1987}.  Similarly, Elyutin and Klyshko have introduced a single-mode cubic form of $\hat{A}$ to describe third subharmonic generation \cite{Elyutin1990}.  More recently, Zelyay, Dey, and Hussin used the term \emph{generalized squeezed states} to describe a generalization of single-mode squeezing in terms of a generalized number operator, from which are defined generalized creation and annihilation operators \cite{Zelaya2018}.  All such generalizations are defined only for single-mode operators and therefore cannot be used to represent multiphoton entangled states.  A general description of higher-order, multimode forms is therefore lacking.

In this paper, we consider the theoretical description of a general multimodal \textit{pinched} vacuum state that may be used to represent general $n$-photon entangled states.  The term arises from the distorted or ``pinched'' appearance of the corresponding Wigner function in phase space \cite{DellAnno2004}.  The new pinching operator is different from previously-considered squeezing operators in that polynomials of creation and annihilation operators may be larger than quadratic, and interactions between modes are therefore described by a tensor.  This description includes coherent states, for $n = 1$, entangled photon pairs, for $n = 2$, as well as other multiphoton entangled states.

In describing the pinching operator we do not consider any specific physical system but, rather, seek to develop a general theoretical framework under an assumption of ideal fabrication and operation conditions involving only a limited number of modes.  In particular, we do not consider imperfections that may arise from, say, phase mismatch, frequency detuning, or pump linewidth.  Similarly we do not consider the effects of pump depletion other than to suppose that the pump may be treated classically and, hence, subsumed in the pinching tensor description.

The outline of the paper is as follows.  In Sec.\ \ref{sec:PinchedStates} we give a general mathematical description of multiphoton pinched vacuum states in terms of a rank-$n$ pinching tensor and relate this to a general $n$-photon entangled Fock state.  Next, in Sec.\ \ref{sec:Bogo} we derive a recursive relation for the Bogoliubov transformations of the corresponding creation and annihilation operators and give expressions for the approximate $n$-photon state.  In Sec.\ \ref{sec:Validation} we provide a validation of these approximations for $n \ge 3$ by simulating certain multiphoton entangled states and examining both violations of Mermin's inequality and state fidelity deduced from quantum state tomography.  Our conclusions are summarized in Sec.\ \ref{sec:Conclusion}.

%%%%%%%%%%%%%%%%%%%%%%%%%%%%%%%%%%%%%%%%%%%%%%%%%

\section{Pinched States}
\label{sec:PinchedStates}
Let $\xi_{i_1,\ldots,i_n} \in \mathbb{C}$ for $i_1,\ldots,i_n \in \{1,\ldots,nd\}$ be the elements of a $n^{\rm th}$-order pinching tensor $\mtx{\xi}$ representing, in the weak pinching limit, $n \ge 1$ photons each in, say, a distinct spatial mode and each in a superposition over $d$ internal degrees of freedom that, together, comprise $nd$ distinct modes.  (The latter may represent orthogonal polarization modes, orbital angular momentum modes, distinct spatial modes, or other degrees of freedom.)  The corresponding pinching operator $\hat{S} = \exp(\hat{A})$ is given by
\begin{equation}
\hat{S} = \exp\left[ \frac{1}{n!} \sum_{i_1,\ldots,i_n} \left( \xi_{i_1,\ldots,i_n} \hat{a}_{i_1}^\dagger \cdots \hat{a}_{i_n}^\dagger - \mathrm{h.c} \right) \right] \; ,
\label{eqn:PinchingOperator}
\end{equation}
where $\hat{a}_i^\dagger$ and $\hat{a}_i$ are the creation and annihilation operators, respectively, for mode $i$ and \mbox{``h.c.''} denotes the Hermitian conjugate of the term to the left.  Since operators corresponding to distinct modes commute, $\xi_{i_1,\ldots,i_n}$ must be a symmetric tensor that is invariant under any permutation of its indices.  Two important special cases are $n=1$, corresponding to the familiar $d$-mode displacement operator, and $n=2$, corresponding to a multimodal Gaussian squeezing operator.

For optical systems, the pinching tensor $\mtx{\xi}$ may be related to the nonlinear optical susceptibility tensor $\mtx{\chi}$ \cite{Boyd2008}.  In particular, a susceptibility tensor $\mtx{\chi}^{(n-1)}$ of order $n-1$ can give rise to a pinching tensor of up to order $n$.  If one or more modes are treated classically, the order of the pinching tensor may be less than that of the of the susceptibility.  For example, a $\mtx{\chi}^{(2)}$ nonlinearity can be used to generate entangled photon pairs through spontaneous parametric down conversion, while a $\mtx{\chi}^{(3)}$ nonlinearity can produce entangled triplets through four-wave mixing \cite{Jain1983}.  More recently, atomic vapors exhibiting a $\chi^{(5)}$ nonlinearity have been used to generate triphoton entangled states via spontanrous six-wave mixing \cite{Li2024}.  With a sufficiently strong higher order susceptibility, an entangled state consisting of any number of photons can, in theory, be produced.

Since $\mtx{\xi}$ is symmetric, the maximum number of unique elements is smaller than the total number, $(nd)^n$, of elements.  Since there are $\binom{nd}{k}$ elements with $k$ distinct indices, the maximum number of distinct elements is
\begin{equation}
D_{n}^{(d)} = \sum_{k=1}^{n} \binom{nd}{k} \; .
\end{equation}
Note that $D_{n}^{(1)} = 2^n-1$ and, it can be shown,
\begin{equation}
D_{n}^{(2)} = \frac{1}{2} \left[ 2^{2n} + \binom{2n}{n} - 2 \right] \ge \binom{2n}{n} \ge 2^n \; .
\end{equation}
Thus, for $d \ge 2$ there are sufficient degrees of freedom to represent any entangled state of $n$ photons.

We are interested in the case of \emph{weak, high-order} pinching to represent multiphoton entangled states.  For a 
modal vacuum state $\ket{\vec{0}}$, the wavefunction of the pinched state is given by $\ket{\Psi} = \hat{S} \ket{\vec{0}}$, so the pinched state, to lowest order in $\mtx{\xi}$, may be written
\begin{equation}
\ket{\Psi} \approx \ket{\vec{0}} + \frac{1}{n!} \sum_{i_1,\ldots,i_n} \xi_{i_1,\ldots,i_n} \hat{a}_{i_1}^\dagger \cdots \hat{a}_{i_n}^\dagger \ket{\vec{0}} \; .
\end{equation}
Squeezed states of this form are commonly used to generate entangled photon pairs through the process of parametric downconversion.  This requires removing the vacuum term and higher order terms through post-selection on specific coincident detection events.  Neglecting the vacuum and higher order terms, the second term represents a general Fock state of $n$ photons distributed among $nd$ modes that may represent, for example, an $n$-photon entangled state.

For greater clarity, the following extended ket notation will be used to describe a Fock state of $nd$ modes with occupation numbers $\vec{N} = (N_1, \ldots, N_{nd})$:
\begin{equation}
\ket{\vec{N}} = \bigket{cccc}{ N_1 & N_{d+1} & \cdots & N_{(n-1)d+1} \\ \vdots & \vdots & \ddots & \vdots \\ N_d & N_{2d} & \cdots &N_{nd} } \; .
\end{equation}
Here, each column corresponds to a different photon, while each row corresponds to an internal degree of freedom, such as polarization.  Thus, a general single-qubit polarization state may be represented by
\begin{equation}
\alpha_0 \ket{H} + \alpha_1 \ket{V} \propto \xi_1 \bigket{c}{ 1 \\ 0 } + \xi_2 \bigket{c}{ 0 \\ 1 } \; , 
\end{equation}
while a general two-qubit state may be represented by
\begin{equation}
\begin{split}
&\alpha_{00} \ket{HH} + \alpha_{01} \ket{HV} + \alpha_{10} \ket{VH} + \alpha_{11} \ket{VV} \\
&\propto \xi_{13} \bigket{cc}{ 1 & \!\!\!1 \\ 0 & \!\!\!0 } + \xi_{14} \bigket{cc}{ 1 & \!\!\!0 \\ 0 & \!\!\!1 } + \xi_{23} \bigket{cc}{ 0 & \!\!\!1 \\ 1 & \!\!\!0 } + \xi_{24} \bigket{cc}{ 0 & \!\!\!0 \\ 1 & \!\!\!1 } \; ,
\end{split}
\end{equation}
with all other tensor elements zero.  Following this general scheme, any $n$-qubit state may be represented by an $n^\mathrm{th}$-order pinching tensor with $2^{n+1}$ nonzero elements.

Although the Hilbert space that can be represented by an $n^\mathrm{th}$-order pinching tensor is vast, only $n$ photons are needed to statistically represent it, to lowest order.  This suggests that entangled states arising from pinching operators may be represented more compactly in the Heisenberg picture.  In the following section, we consider the representation of pinched states in terms of the pinched creation and annihilation operators obtained via the Bogoliubov transformations.  This operator representation will also faciliate the representation of the state in terms of complex random vectors.

%%%%%%%%%%%%%%%%%%%%%%%%%%%%%%%%%%%%%%%%%%%%%%%%%

\section{Bogoliubov Transformations}
\label{sec:Bogo}

The Bogoliubov transformation describes how creation and annihilation operators transform under the action of the pinching operator.  If $\hat{a}_i$ is the annihilation operator for mode $i$, then the corresponding pinched annihilation operator is
\begin{equation}
\hat{b}_i = \hat{S}^\dagger \hat{a}_i \hat{S} = e^{-\hat{A}} \hat{a}_i e^{\hat{A}} \; ,
\end{equation}
since $\bra{\Psi} \hat{a}_i \ket{\Psi} = \bra{\vec{0}} \hat{S}^\dagger\hat{a}_i\hat{S} \ket{\vec{0}} = \bra{\vec{0}} \hat{b}_i \ket{\vec{0}}$.

Using the fact that $\hat{A}$ is skew Hermitian, we prove in Appendix \ref{app:Agnes} the following general result:
\begin{equation}
\hat{b}_i = \sum_{k=0}^{\infty} \frac{\hat{C}_i^{(k)}}{k!} \; ,
\label{eqn:Bogoliubov}
\end{equation}
where $\hat{C}_i^{(k)}$ is defined recursively by
\begin{equation}
\begin{split}
\hat{C}_i^{(0)} &= \hat{a}_i \\
\hat{C}_i^{(k)} &= [\hat{C}_i^{(k-1)}, \, \hat{A}]
\end{split}
\end{equation}
and given more explicitly by
\begin{equation}
\hat{C}_i^{(k)} = \sum_{\ell=0}^{k} \binom{k}{\ell} (-\hat{A})^{k-\ell} \hat{a}_i \hat{A}^{\ell} \; .
\label{eqn:Ck}
\end{equation}
Analytic expressions for $\hat{b}_i$ are known for the cases $n = 1$ and $n = 2$, corresponding to displacement and squeezing, respectively, as illustrated below.  Our primary focus will be on cases in which $n \ge 3$.

%----------------------------------------------------------------------------------------------------------------------------------

\subsection{Case $n=1$: Displacement Operators}

When $\mtx{\xi}$ is a first-order tensor (i.e., a vector), the pinching operator takes the simple form
\begin{equation}
\hat{S} = \exp\left[\sum_{j=1}^{d} \left( 
\xi_{j} \hat{a}_{j}^\dagger -
\xi_{j}^* \hat{a}_{j} \right) \right] \; ,
\label{eqn:DisplacementOperator}
\end{equation}
which we recognize as the displacement operator for $d$ modes.  Using the bosonic commutation relation
\begin{equation}
[\hat{a}_i, \hat{a}_j^\dagger] = \delta_{ij} \hat{1} \; ,
\end{equation}
we find that $\hat{C}_i^{(1)} = \xi_i \hat{1}$ and $\hat{C}^{(k)} = \hat{0}$ for $k \ge 2$.  Thus,
\begin{equation}
\hat{b}_i = \hat{a}_i + \xi_i \hat{1} \; .
\end{equation}

%----------------------------------------------------------------------------------------------------------------------------------

\subsection{Case $n = 2$: Gaussian Squeezing Operators}

When $\mtx{\xi}$ is a second-order tensor (i.e., a matrix), the pinching operator takes the form
\begin{equation}
\hat{S} = \exp\left[ \frac{1}{2} \sum_{i,j=1}^{2d} \left( 
\xi_{ij} \hat{a}_{i}^\dagger \hat{a}_{j}^\dagger -
\xi_{ij}^* \hat{a}_{i} \hat{a}_{j} \right) \right] \; ,
\label{eqn:SqueezingOperator}
\end{equation}
which we recognize as a multimode squeezing operator.  Let $\hat{\vec{a}}^\dagger = [\hat{a}_1^\dagger, \ldots, \hat{a}_{2d}^\dagger]^\mathsf{T}$ and $\hat{\vec{a}} = [\hat{a}_1, \ldots, \hat{a}_{2d}]^\mathsf{T}$ denote column vectors of the creation and annihilation operators, respectively.  The vector $\hat{\vec{b}}$ is defined similarly.  In Appendix \ref{app:Angus} we derive the familiar result
\begin{equation}
\vec{b} = (\cosh\mtx{R}) \hat{\vec{a}} + (\sinh\mtx{R}) \mtx{Q} \hat{\vec{a}}^\dagger \; ,
\label{eqn:Damien}
\end{equation}
where $\mtx{\xi} = \mtx{R} \mtx{Q}$ is a polar decomposition of $\mtx{\xi}$ with positive semi-definite matrix $\mtx{R}$ and unitary matrix $\mtx{Q}$.

%----------------------------------------------------------------------------------------------------------------------------------

\subsection{Approximations for $n \ge 3$}

The general case $n \ge 3$ does not lend itself to a simple, closed-form solution, but we may use Eqn.\ (\ref{eqn:Bogoliubov}) as a scheme for systematically approximating $\hat{b}_i$.  The result, to second order, is given in Appendix \ref{app:Beatrice}.

As a specific example, consider the pinching operator for $n=3$, given by
\begin{equation}
\hat{S} = \exp\left[\frac{1}{6} \sum_{i,j,k} \left( 
\xi_{ijk} \hat{a}_{i}^\dagger \hat{a}_{j}^\dagger \hat{a}_{k}^\dagger -
\mathrm{h.c.} \right)\right] \; .
\label{3photonSqueezingOperator}
\end{equation}
For this case we may write $\hat{b}_i$ to second order as
\begin{equation}
\hat{b}_i = \hat{a}_i + \hat{C}_i^{(1)} + \frac{1}{2} \hat{C}_i^{(2)} + \cdots \; ,
\end{equation}
where
\begin{equation}
\hat{C}^{(1)}_i = \frac{1}{2} \sum_{j,k} \xi_{ijk} \hat{a}^\dagger_j \hat{a}^\dagger_k
\end{equation}
and
\begin{equation}
\hat{C}^{(2)}_i = \frac{1}{4} \sum_{j,k} \xi_{ijk} \sum_{j',k'} \xi_{jj'k'} \left( \hat{a}_{j'} \hat{a}_{k'} \hat{a}_{k}^\dagger + \hat{a}_{k}^\dagger \hat{a}_{j'} \hat{a}_{k'} \right) \; .
\end{equation}
Rewriting the latter in normal order, we have
\begin{equation}
\hat{C}^{(2)}_i = \frac{1}{2} \sum_{j,k,\ell} \xi_{ijk} \xi^*_{jk\ell} \hat{a}_{\ell} +\frac{1}{2} \sum_{j,k,\ell,m} \xi_{ijk} \xi^*_{k{\ell}m} \hat{a}^\dagger_j \hat{a}_{\ell} \hat{a}_m \; .
\end{equation}

Note that the first order approximation is both normally and symmetrically ordered.  In particular, the symmetric ordering allows for a positive Wigner function representation in terms of complex random variables, which we shall make use of in Sec.\ \ref{sec:Validation}.

%----------------------------------------------------------------------------------------------------------------------------------

\subsection{Low-Order Approximations}

As we have seen, when $\mtx{\xi}$ is a tensor of rank three or higher, $\hat{C}_i^{(k)}$ does not have a simple closed form, as the commutators of $\hat{a}_i$ with more than two creation operators will have two or more creation operators. Repeated commutations will therefore lead to larger products of creation and annihilation operators in each term. This can be a significant impediment to understanding cubic and higher-order pinching operators.

However, in the present context we are only interested in low-order approximations, as these are what are needed to approximate entangled states.  Including higher-order terms would yield a better approximation to the pinched state but a worse approximation to the $n$-photon entangled state.  Higher-order terms may, however, be needed to more faithfully model experiments that involve high-intensity pumping or materials with strong, high-order nonlinearities.  For our purposes, however, low-order approximations are actually preferable.

We may consider these approximations more formally by defining the order-$p$ pinching operator
\begin{equation}
\hat{S}^{(p)} = \sum_{k=0}^{p} \frac{\hat{A}^k}{k!} \; .
\end{equation}
Although $\hat{S}^{(p)}$ is not a unitary operator, it may act upon the vacuum state $\ket{\vec{0}}$ to produce an unnormalized approximation $\ket{\Psi^{(p)}} = \hat{S}^{(p)} \ket{\vec{0}}$ to the pinched state $\ket{\Psi} = \hat{S} \ket{\vec{0}}$.  In particular, $\ket{\Psi^{(p)}} - \ket{\vec{0}} \propto \ket{\psi}$ provides an unnormalized approximation to the $n$-photon Fock state $\ket{\psi}$.  We expect that by making $\mtx{\xi}$ small and removing the vacuum term (e.g., by post-selection of photon detection events), we may obtain a good approximation of $\ket{\psi}$.

In Appendix \ref{app:Charlene}, we show that $\hat{\vec{b}}^{(p)}$ (in the Heisenberg picture) may be used to approximate $\ket{\Psi^{(p)}}$ (in the Schr\"{o}dinger picture).  In particular, $\hat{\vec{b}}^{(1)}$ may be used as a means of approximating $\ket{\psi}$, in superposition with the vacuum state.  This will be used in the following section for performing numerical simulations of multiphoton entangled states.

%%%%%%%%%%%%%%%%%%%%%%%%%%%%%%%%%%%%%%%%%%%%%%%%%

\section{Classical Simulation}
\label{sec:Validation}

Having determined a low-order approximation for general pinched states, we now consider how these results may be used for a fully classical simulation.  In this section we consider a method for numerically sampling from a distribution corresponding to a weakly pinched state and examine whether this method can be used to accurately represent multiphoton entangled states after post-selection.

Non-Gaussian pure states, such as pinched vacuum states, are known to be computationally inefficient to simulate.  An efficient classical scheme can therefore only be possible for mixed states.  Prior work has shown that efficient classical simulations of non-Gaussian states with positive Wigner functions are indeed possible, however, with the restriction that one consider only Gaussian measurements \cite{Veitch2013}.  This work considers simulations of more general non-Gaussian measurements, in particular projective measurements.  Furthermore, as we will show, the proposed method scales not only polynomially but \emph{linearly} with the number of modes in generating random realizations that are representative of the pinched vacuum state in the weak pinching regime.

Using the pinched vacuum states one can go further in simulating, at least approximately, pure multiphoton entangled states through post-selection.  Of course, pure states are themselves an idealization, and any experimentally prepared non-Gaussian state will inevitably be mixed.  For this reason we seek a classical simulation scheme capable of producing mixed states that, after post-selection, provide comparable or better fidelity, as conventionally assessed by measurement, to that which has been experimentally demonstrated.  For this reason, we will consider several examples comparing our simulation results to similar experimentally obtained results.

Prior work has demonstrated that classical simulations with weakly squeezed Gaussian entangled states, combined with amplitude threshold detectors, can successfully emulate two-photon entangled states to a high degree of fidelity, albeit with limited detection efficiency \cite{LaCour2021}.  In this section we explore whether this approach can be extended to more general multiphoton entangled states using weakly pinched non-Gaussian states approximated to first order.  To do so, we must first make a connection between the quantum operator representation and representative classical random variables.

The optical equivalence theorem states that the expectation value of quantum operators in a certain ordering will be given by the analogous expectation value of the corresponding classical observables with respect to a certain quasi-probability density function \cite{Sudarshan1963,Cahill1969}.  The Wigner distribution is one such function that may be used for symmetrically ordered operators and, if everywhere positive, may be considered a true probability density function.  The Wigner distribution of a multimode squeezed vacuum state, for example, is Gaussian; therefore, replacing each annihilation operator $\hat{a}_i$ in Eqn.\ (\ref{eqn:Damien}) with an independent complex Gaussian random variable $a_i$ having a mean of zero and a variance of one half, corresponding to the modal vacuum energy $\frac{1}{2} \hbar \omega_i$, gives a statistically equivalent description for symmetrically ordered operators.  Physically, this may be viewed as treating the vacuum modes as classical, stochastic fields interacting with a nonlinear material \cite{Kulkarni2022}.

For pinching tensors of rank greater than two, the pinched vacuum state is non-Gaussian and, since it is pure, the Wigner function must be negative, so an exact representation in terms of random variables is impossible.  If the pinched state is not pure, however, the Wigner distribution of the corresponding mixed state may still be positive.  This means that an equivalent random variable description may be possible.  Simply replacing $\hat{a}_i$ with $a_i$ in Eqn.\ (\ref{eqn:PinchingOperator}) will not suffice, as this reduces $\hat{S}$ to the identity.  We note, however, that the first-order approximation $\hat{b}_i^{(1)}$ of the Bogoliubov transformed operators is, in fact, symmetrically ordered, suggesting that a random variable representation may be possible in this regime.  Substituting each $\hat{a}_i$ in Eqn.\ (\ref{eqn:bp}) with a random variable $a_i$ as before, we obtain a new random variable $b_i$ that is well defined but non-Gaussian due to the product of operators (treated now as random variables) appearing in $\hat{C}^{(1)}$.  Doing so creates the equivalent of a non-Gaussian mixed state \cite{Walschaers2021}.  The remaining question we now address is whether the fundamentally quantum nature of the original pinched state is lost in this approximation and to what degree it may be used to approximate a multiphoton entangled state.

In what follows, we will use the approach described above to define a set of random variables that, we hypothesize, can provide an approximation of the mixed pinched state valid in the weak pinching regime.  We will then use these non-Gaussian random variables to model an entangled multiphoton Fock state by performing post-selection to remove the vacuum term.  This is done using a combination of local transformations and a threshold detector model to achieve discrete measurement outcomes.  Combining the results of various measurement bases, we then perform full state tomography and thereby obtain an inferred density matrix that is compared against the desired multiphoton entangled state.  As a further assessment of the degree of entanglement preserved in these approximations, we consider violations of the Mermin inequality for certain multiphoton states.

%---------------------------------------------------------------------------------------------------------------------

\subsection{State Generation}

Consider a weakly pinched multimode vacuum state approximating a Greenberger-Horne-Zeilinger (GHZ) state.  To represent this state for $n=3$ and $d=2$ we use a pinching operator of the form described in Eqn.\ (\ref{3photonSqueezingOperator}) and define a pinching tensor whose elements are given by
\begin{equation}
\xi_{ijk} = \begin{cases}
            r & \text{if }(i,j,k)\in\text{Perm}(1,3,5) \; , \\
            r e^{i\theta} & \text{if }(i,j,k)\in\text{Perm}(2,4,6) \; , \\
            0 & \text{otherwise} \; ,
            \end{cases}
\label{xiGHZ}
\end{equation}
where $r$ is the pinching strength, $\theta \in \mathbb{R}$ is a relative phase angle, and $\mathrm{Perm}(1,3,5)$, say, is the set of six permutations of $(1,3,5)$.  The pinching tensor above provides an approximate pinched annihilation operator
\begin{equation}
\hat{b}_i \approx \hat{a}_i+\frac{1}{2}\ \sum_{j=1}^{6} \sum_{k=1}^{6} \xi_{ijk} \hat{a}_j^\dagger \hat{a}_k^\dagger
\label{b_approx}
\end{equation}
and pinched state $\ket{\Psi} \approx \ket{\vec{0}} + \sqrt{2} \, r \ket{\text{GHZ}}$, where
\begin{equation}
\ket{\text{GHZ}} = \frac{1}{\sqrt{2}} \Bigl[ \ket{HHH} + e^{i\theta} \ket{VVV} \Bigr] \; .
\end{equation}

If we denote the six spatial and polarization modes $1, 2, 3, 4, 5, 6$ by the symbols $1H, 1V, 2H, 2V, 3H, 3V$, respectively, then each $b_i$ is given explicitly by
\begin{equation}
\begin{pmatrix} b_{1H} \\ b_{1V} \\ \vspace{-0.5em} \\ b_{2H} \\ b_{2V} \\ \vspace{-0.5em} \\ b_{3H} \\ b_{3V} \end{pmatrix}
=
\begin{pmatrix} a_{1H} \\ a_{1V} \\ \vspace{-0.5em} \\ a_{2H} \\ a_{2V} \\ \vspace{-0.5em} \\ a_{3H} \\ a_{3V} \end{pmatrix}
+
r
\begin{pmatrix}
a_{2H}^* a_{3H}^* \\
e^{i\theta} a_{2V}^* a_{3V}^* \\
\vspace{-0.5em} \\
a_{3H}^* a_{1H}^* \\
e^{i\theta} a_{3V}^* a_{1V}^* \\
\vspace{-0.5em} \\
a_{1H}^* a_{2H}^* \\
e^{i\theta} a_{1V}^* a_{2V}^*
\end{pmatrix} \; .
\label{eqn:GHZ3}
\end{equation}

Similarly, a maximally entangled W state of the form
\begin{equation}
\ket{\text{W}} = \frac{1}{\sqrt{3}} \Bigl[ \ket{HHV} + e^{i\theta_1} \ket{HVH} + e^{i\theta_2} \ket{VHH} \Bigr]
\end{equation}
may be represented by the random vector
\begin{equation}
% \begin{pmatrix} b_{1H} \\ b_{1V} \\ \vspace{-0.5em} \\ b_{2H} \\ b_{2V} \\ \vspace{-0.5em} \\ b_{3H} \\ b_{3V} \end{pmatrix}
 % =
\begin{pmatrix} a_{1H} \\ a_{1V} \\ \vspace{-0.5em} \\ a_{2H} \\ a_{2V} \\ \vspace{-0.5em} \\ a_{3H} \\ a_{3V} \end{pmatrix}
+
\sqrt{\frac{2}{3}} \, r
\begin{pmatrix}
a_{2H}^* a_{3V}^* + e^{i\theta_1} a_{2V}^* a_{3H}^* \\
e^{i\theta_2} a_{2H}^* a_{3H}^* \\
\vspace{-0.5em} \\
a_{3V}^* a_{1H}^* + e^{i\theta_2} a_{3H}^* a_{1V}^* \\
e^{i\theta_1} a_{3H}^* a_{1H}^* \\
\vspace{-0.5em} \\
e^{i\theta_1} a_{1H}^* a_{2V}^* + e^{i\theta_2} a_{1V}^* a_{2H}^* \\
a_{1H}^* a_{2H}^* \\
\end{pmatrix} \; .
\label{eqn:W3}
\end{equation}
Note that Eqns.\ (\ref{eqn:GHZ3}) and (\ref{eqn:W3}) use nonlinear combinations of Gaussian random variables and, for $r \neq 0$, yield non-Gaussian random variables.  In accordance with Hudson's theorem \cite{Hudson1974}, the corresponding probability density function represents a positive Wigner function for a mixed quantum state.

Generalizing these results, an $n$-photon GHZ state
\begin{equation}
\ket{\text{GHZ}} = \frac{1}{\sqrt{2}} \left[ \; \bigotimes_{k=0}^{n-1} \ket{H}_k + e^{i\theta} \bigotimes_{k=0}^{n-1} \ket{V}_k \right] \; ,
\end{equation}
may be approximated with the random variables
\begin{subequations}
\begin{align}
b_{iH} &= a_{iH} + r \prod_{j \neq i} a_{jV}^* \; , \\
b_{iV} &= a_{iV} + r e^{i\theta} \prod_{j \neq i} a_{jH}^* \; ,
\end{align}
\end{subequations}
while an $n$-photon W state
\begin{equation}
\ket{\text{W}} = \frac{1}{\sqrt{n}} \sum_{j=0}^{n-1} e^{i\theta_j} \ket{V}_j \bigotimes_{k \neq j} \ket{H}_k 
\end{equation}
may be approximated by the random variables
\begin{subequations}
\begin{align}
b_{iH} &= a_{iH} + \sqrt{\frac{2}{n}} \, r \sum_{j \neq i} e^{i\theta_{n-j}} a_{jV}^* \prod_{k \not\in \{i, j\}} a_{kH}^* \; , \\
b_{iV} &= a_{iV} + e^{i\theta_{n-i}} \prod_{j \neq i} a_{jH}^* \; .
\end{align}
\end{subequations}

As continuous variables, it is possible to verify nonseparability directly using moment-based entanglement witnesses \cite{Zhang2023}.  For modeling discrete-variable states; however, some form of binary detection is needed.  The detectors themselves are modeled such that a detection occurs when the incident amplitude exceeds a given threshold $\gamma$. For example, a result of $H$ is obtained for photon $i$ if $|b_{iH}| > \gamma$ and $|b_{iV}| \le \gamma$.  Conversely, a result of $V$ is obtained if $i$ if $|b_{iV}| > \gamma$ and $|b_{iH}| \le \gamma$.  Measurements in different bases may be performed by first applying the appropriate unitary transformation to the Jones vector $[b_{iH}, b_{iV}]^\mathsf{T}$.  Thus, measurements are treated as local and deterministic, given the particular realization for the incident light.  Modeling detectors in this manner was first proposed by Khrennikov as a means of exploring the quantum-classical boundary \cite{Khrennikov} and has since been used to describe various observed statistical behavior of both coherent and entangled quantum light \cite{LaCour2020,LaCour2021}.  Although this stochastic model does not create photons as such, each random realization may be thought to correspond to a single coherence time.  The detector model, then, corresponds to a coincidence window equal to this coherence time, with no subsequent dead time.  Note that, even when the pinching tensor is zero, detections may still occur under this model.  These ``dark counts'' may be reduced by increasing the threshold, but they cannot be eliminated altogether.

For the pinched GHZ and W states, we generated a large number of realizations with different pinching strengths $r$ and detection thresholds $\gamma$.  We then performed quantum state tomography (QST) and used the resulting density matrices to compute the fidelity of the inferred states.  Finally, the generated states were used to investigate possible violations of the Mermin inequality.  The results of these numerical simulations are described in the following sections.

%---------------------------------------------------------------------------------------------------------------------

\subsection{Quantum State Tomography}

\begin{figure}[ht]
\ifshowfigures
\includegraphics[width=\columnwidth]{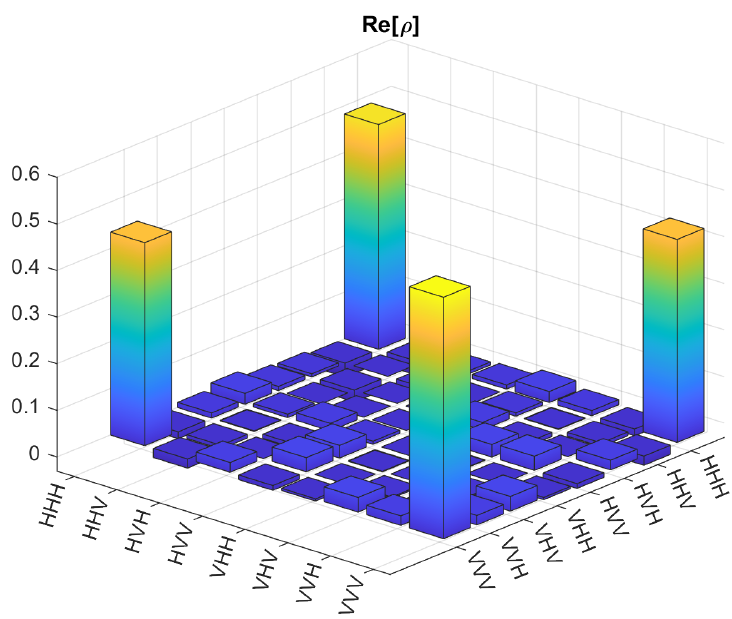}
\centerline{(a)}
\vspace{0.5em}

\includegraphics[width=\columnwidth]{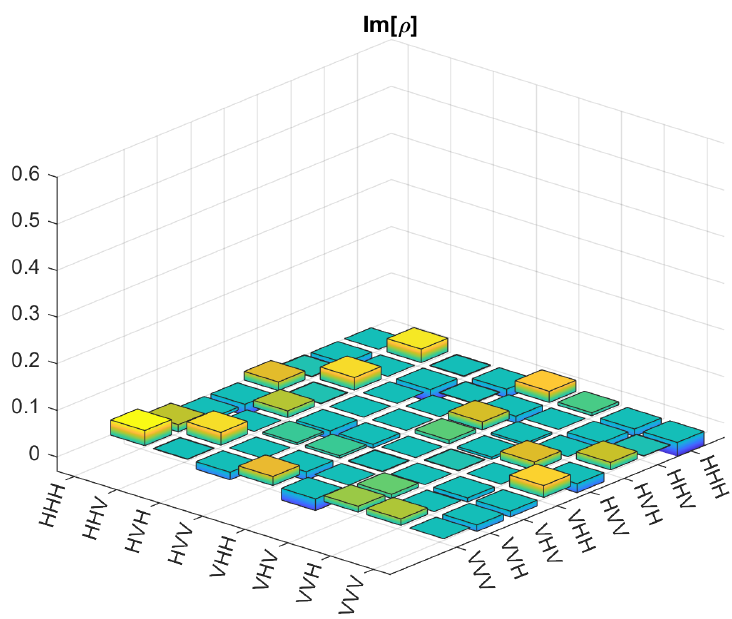}
\centerline{(b)}
%\begin{subfigure}[b]{\columnwidth}
%\includegraphics[width=\columnwidth]{Figs-GHZ3cityscapeRe}
%\caption{$\mathrm{Re}[\boldsymbol{\rho}]$}
%\label{fig:GHZ3cityscapeRe}
%\end{subfigure}
%\begin{subfigure}[b]{\columnwidth}
%\includegraphics[width=\columnwidth]{Figs-GHZ3cityscapeIm}
%\caption{$\mathrm{Im}[\boldsymbol{\rho}]$}
%\label{fig:GHZ3cityscapeIm}
%\end{subfigure}
\fi
\caption{(Color online) Density matrix $\mtx{\rho}$ estimated from QST for a simulated pinched three-photon GHZ state with $\theta = 0$, $r=0.6$, $\gamma=2.0$, and a fidelity of $0.93 \pm 0.01$.  The top subplot (a) shows the real part of $\mtx{\rho}$, and the bottom subplot (b) shows the imaginary part.}
\label{fig:GHZ3cityscape}
\end{figure}

We prepared both GHZ and W states and performed dual-mode linear quantum state tomography (QST) on each using different values of $r$ and $\gamma$.  Ideal GHZ and W states are non-Gaussian entangled pure states, so our preparations can only provide an approximation.  We use fidelity to these pure states, as deduced from QST, as our primary metric of quality.  To perform QST, we measured in all $4^n$ Pauli basis states using local operations and the aforementioned detection model over a sample of about a million ($2^{20}$) random realizations.  State fidelity, $F$, was computed from the inferred density matrix, $\mtx{\rho}$, using $F = \bra{\rm GHZ}\mtx{\rho}\ket{\rm GHZ}$.  For certain parameter values we found that the approximate pinching operator created a high-fidelity GHZ state.  An example of an estimated density matrix for a pinched three-photon GHZ state exhibiting a fidelity of $0.93 \pm 0.01$ is shown in Fig. \ref{fig:GHZ3cityscape}.  For comparison, three-photon GHZ states have been generated experimentally with fidelities only as high as $0.82 \pm 0.03$ \cite{Park2021}.

\begin{figure}[ht]
\ifshowfigures
\includegraphics[width=\columnwidth]{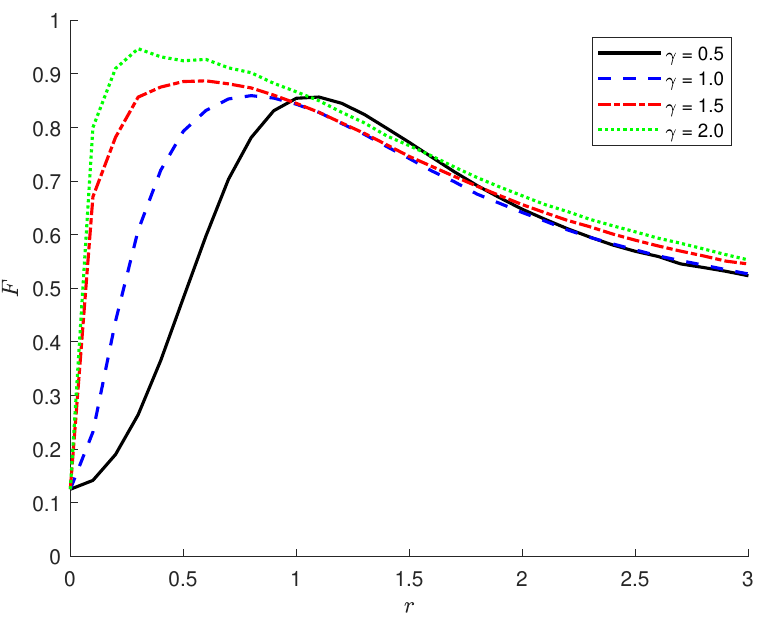}
\fi
\caption{(Color online) Plot of fidelity $F$ for an approximate three-photon GHZ state versus pinching strength $r$ and detection thresholds $\gamma = 0.5$ (black solid line), $\gamma = 1.0$ (blue dashed line), $\gamma = 1.5$ (red dash-dot line), and $\gamma = 2.0$ (green dotted line).}
\label{fig:GHZ3fidelity}
\end{figure}

Varying the pinching strength, $r$, or the detection threshold, $\gamma$, can yield different QST fidelities.  For the above three-photon GHZ state we varied $r$ from 0.0 to 3.0 and considered $\gamma$ values of 0.5, 1.0, 1.5, and 2.0.  The results are shown in Fig.\ \ref{fig:GHZ3fidelity}.  We see that for each threshold there is an optimum value of $r$ that yields the best fidelity.  If $r$ is too small, the pinched state is dominated by the vacuum terms and has low fidelity.  Similarly, if $r$ is too large then the $n$-photon approximation breaks down.  Higher threshold values give lower dark counts but also lower detection counts.  In some regimes, such as a high threshold and low pinching strength, the number of valid detections is too low to sample adequately.  Interestingly, we also note that, as detection threshold increases, the optimum fidelity increases as well and occurs at lower pinching strengths.  Note that large values of $r$ violate the small pinching strength assumption, and, as a result, the fidelity degrades systematically.

\begin{figure}[ht]
\ifshowfigures
\includegraphics[width=\columnwidth]{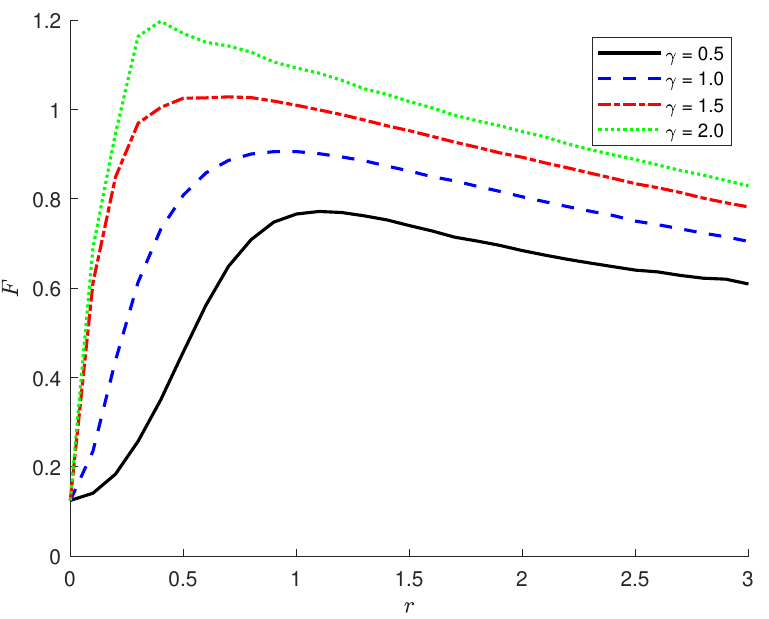}
\fi
\caption{(Color online) Plot of fidelity $F$ for an approximate three-photon W state versus pinching strength $r$ and detection thresholds $\gamma = 0.5$ (black solid line), $\gamma = 1.0$ (blue dashed line), $\gamma = 1.5$ (red dash-dot line), and $\gamma = 2.0$ (green dotted line).  The over-unity results for $\gamma = 1.5$ and $\gamma = 2.0$ are a numerical artifact of linear QST.}
\label{fig:W3fidelity}
\end{figure}

A similar analysis was performed for the three-photon W state (with $\theta_1 = \theta_2 = 0$); the results are shown in Fig.\ \ref{fig:W3fidelity}.  We find qualitatively similar behavior to that of the three-photon GHZ state, although the fidelity for the W state tends to be higher than that of the GHZ state for the same parameter settings.  We also note that for large threshold values the optimal fidelity can actually go above unity.  This is a well-known numerical artifact of linear QST for high-fidelity states and can be removed by using a more sophisticated maximum likelihood approach with a positive semi-definite constraint \cite{Altepeter2005}.  For comparison, photonic experiments have demonstrated three-photon W-state fidelities as high as $0.978 \pm 0.008$ \cite{Zhang2016}.

Finally, we considered fidelity as a function of photon number, $n$, for a pinched GHZ state with $\theta = 0$.  For $n = 2, 3, 4, 5, 6$ we performed QST with fixed parameter values of $r = 0.6$ and $\gamma = 2.0$ and obtained fidelities of $F = 0.98 \pm 0.01, \, 0.93 \pm 0.01, \, 0.75 \pm 0.02, \, 0.61 \pm 0.02, \, 0.56$, respectively.  Comparing to experiment, four-photon and six-photon GHZ states have been generated optically with fidelities as high as $0.833 \pm 0.004$ and $0.710 \pm 0.016$, respectively \cite{PanGroup2016}.  More recently, GHZ states with 10 and 12 photons have also been generated, fidelities of $0.573 \pm 0.023$ and $0.572 \pm 0.024$, respectively \cite{PanGroup2016,PanGroup2018}.  We expect that our model could achieve higher fidelities using a combination of lower values of $r$, higher values of $\gamma$, and much larger samples.

%---------------------------------------------------------------------------------------------------------------------

\subsection{Mermin Inequality Tests}

The Mermin inequalities have been used to verify multipartite entanglement and test quantum nonlocality \cite{Erven2014}.  Here we shall use them to assess the accuracy of our numerical model.  To define the inequality, we begin by defining the Mermin operator $\hat{M}_n$ for an $n$-photon system, which is given by \cite{Mermin1990}
\begin{equation}
\hat{M}_n = \frac{1}{2} \left[ (\hat{\sigma}_x + i\hat{\sigma}_y)^{\otimes n} + (\hat{\sigma}_x - i\hat{\sigma}_y)^{\otimes n} \right]  \; ,
\end{equation}
where $\hat{\sigma}_x$ and $\hat{\sigma}_y$ are Pauli spin operators.  In particular, $\hat{M}_1 = \hat{\sigma}_x$, $\hat{M}_2 = \hat{\sigma}_x \hat{\sigma}_x - \hat{\sigma}_y \hat{\sigma}_y$, and
\begin{equation}
\hat{M}_3 = \hat{\sigma}_x \hat{\sigma}_x \hat{\sigma}_x - \hat{\sigma}_x \hat{\sigma}_y \hat{\sigma}_y - \hat{\sigma}_y \hat{\sigma}_x \hat{\sigma}_y - \hat{\sigma}_y \hat{\sigma}_y \hat{\sigma}_x \; .
\end{equation}

The maximum expectation value of $\hat{M}_n$ is achieved for the standard GHZ state, $\ket{\text{GHZ}} = \frac{1}{\sqrt{2}} [ \ket{H \cdots H} + \ket{V \cdots V} ]$, and is given by
\begin{equation}
\bra{\text{GHZ}} \hat{M}_n \ket{\text{GHZ}} = 2^{n-1} \; .
\end{equation}
The general form of $\hat{M}_n$ consists of $2^{n-1}$ terms, each of which has eigenvalues of $\pm1$ and consists of an even number of $\hat{\sigma}_y$ spin operators in all combinations of order.  Each term with an odd number of $\hat{\sigma}_y$ pairs has a minus sign and an expectation of $-1$.  All others have a plus sign and an expectation of $+1$.  For example,
\begin{equation}
\hat{M}_4 = \hat{\sigma}_x \hat{\sigma}_x \hat{\sigma}_x \hat{\sigma}_x - ( \hat{\sigma}_x \hat{\sigma}_x \hat{\sigma}_y \hat{\sigma}_y + \cdots ) + \hat{\sigma}_y \hat{\sigma}_y \hat{\sigma}_y \hat{\sigma}_y \; .
\end{equation}

Now, the Mermin inequality states that, for a noncontextual hidden-variable model in which $M_n$ denotes the random variable corresponding to $\hat{M}_n$, we have the following constraint on its expectation value:
\begin{equation}
\mathsf{E}[M_n] \le 
\begin{cases}
2^{n/2} & \text{if $n$ is even} \\
2^{(n-1)/2} & \text{if $n$ is odd.}
\end{cases}
\end{equation}
In what follows, we compute the Mermin statistic using our numerical model for different photon number GHZ states.  Since we post-select on coincident detection events, the noncontextuality condition need not hold, as the coincident events for each measurement setting correspond to a different subensemble.  This allows for a possible violation of the Mermin inequality \cite{Larsson1998}.

\begin{figure}[ht]
\ifshowfigures
\includegraphics[width=\columnwidth]{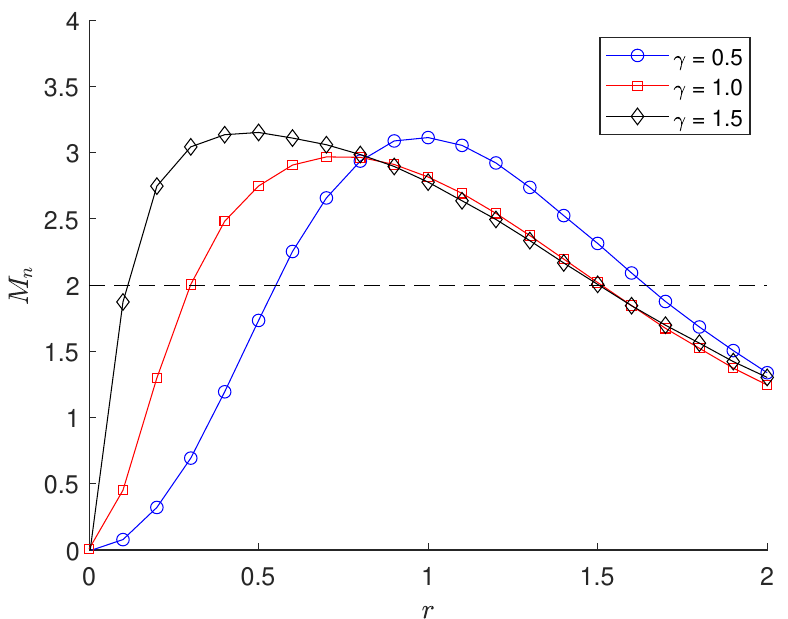}
\fi
\caption{(Color online) Plot of Mermin statistic $M_3$ as a function of pinching strength $r$ for a detection threshold of $\gamma = 0.5$ (red squares), $\gamma = 1.0$ (blue circles), and $\gamma = 1.5$ (black diamonds).  The dashed line is the upper bound of two for noncontextual models.  The quantum upper bound is four.}
\label{fig:Mermin3}
\end{figure}

In Fig.\ \ref{fig:Mermin3} we show the results of measuring the Mermin statistic for an approximate three-photon GHZ state.  The pinching strength, $r$, was varied from 0.0 to 2.0, and two different detection thresholds, $\gamma = 0.5$ and $\gamma = 1.0$, were considered.  We found violations of the Mermin inequality for intermediate values of $r$, with higher values of $\gamma$ necessitating lower values of $r$.  For $\gamma = 0.5$, violations were found for $0.55 < r < 1.64$, and a peak violation of $M_3 = 3.11$ was found for $r = 1.0$.  For $\gamma = 1.0$, violations were found for $0.3 < r < 1.51$, and a peak violation of $M_3 = 2.97$ was found for $r = 0.7$.  Interestingly, the pinching strengths required for maximum violation are larger than those needed for maximum fidelity.  (See Fig.\ \ref{fig:GHZ3fidelity}.)  We believe this is due to the larger number of observables needed to perform QST for computing fidelity, resulting in more stringent demands on the suppression of higher-order Fock states.

Higher thresholds and lower pinching strengths can give stronger violations, but the optimal values will depend on $n$.  We considered violations of Mermin's inequality for different photon number states, with a single value of $r$ and $\gamma$ selected for each value of $n$ that gives near-optimal violation.  A summary of these results for $n \in \{ 3, 4, 5 \}$ is shown in Table \ref{tbl:Mermin}.  The Mermin statistic, $M_n$, was computed over $2^{20}$ random realizations, and this process was repeated twenty times to further assess variability.  For example, for $n = 3$ we chose $r = 0.6$ and $\gamma = 2.3$ and obtained a sample mean of $M_3 = 3.60$ with a standard deviation of $0.03$.   In the table, these results are compared against experimentally observed Mermin inequality violations in photonic and superconducting qubit experiments.  The values of $n$ are common to both our model and the corresponding experiments; however, $r$ and $\gamma$ are model parameters that have no corresponding experimental values.  We find that our model is capable of producing violations that are comparable to or larger than those observed experimentally.  Given that the optical experiments used fusion gates to create the state, it may be difficult to use our work to recommend specific improvements other than to say that the optimal amount of squeezing used in spontaneous parametric down-conversion should be sensitive to the detectors used as well as the complete optical setup.

\begin{table}[ht]
\begin{tabular}{cccccc}
\hline
$n$ \; & $r$ & $\gamma$ & $M_n$ (model) & $M_n$ (exp.) & Ref. \\
\hline
3 \; & 0.6 & 2.3 & \; $3.60 \pm 0.03$ \; & \; $3.48 \pm 0.16$ \; & \cite{Pan2000} \\
4 \; & 0.9 & 0.6 & \; $4.54 \pm 0.02$ \; & \; $4.81 \pm 0.06$ \; & \cite{Alsina2016a} \\
5 \; & 0.5 & 1.0 & \; $5.27 \pm 0.01$ \; & \; $4.05 \pm 0.06$ \; & \cite{Alsina2016a} \\
\hline
\end{tabular}
\caption{Table of Mermin statistics for different photon numbers, $n$.  Here, $r$ and $\gamma$ are the pinching strength and detection threshold, respectively.  The fourth column shows the Mermin statistic for our model, while the fifth column shows the experimentally observed value from the reference in the last column.}
\label{tbl:Mermin}
\end{table}

%%%%%%%%%%%%%%%%%%%%%%%%%%%%%%%%%%%%%%%%%%%%%%%%%

\section{Conclusion}
\label{sec:Conclusion}

We have introduced the pinching operator as a generalization of the multimode displacement and squeezing operators to describe multiphoton entangled states.  The operator is specified by a fully symmetric tensor (i.e., one that is invariant under all permutations of its indices), with the rank corresponding to the number of photons to be represented.  Using the Bogoluibov transformations, we provided a recursive procedure for computing the pinched creation and annihilation operators to any order of approximation.  Truncating the resulting infinite series gives an approximation to the pinched vacuum state, and, with suitably chosen pinching tensors, the resulting non-Gaussian state can be used to approximate an arbitrary multiphoton entangled state.  Explicit forms for generating GHZ and W states were provided.

We then described a procedure for classically simulating multiphoton entangled states using threshold detection and the post-selection of coincident detection events to effect non-Gaussian measurements.  This model was found to be capable of producing high-fidelity simulations of entangled multiphoton experiments, in some cases exceeding the fidelity of photonic states generated experimentally, in a manner that scales linearly with the number of modes.  Fidelity was found to vary with the pinching strength and detector threshold of the model, with intermediate values of pinching strength giving optimal fidelity.  The right combination of low pinching strength and high detection threshold was found to give the best fidelity, but this regime may be difficult to sample numerically.  Model fidelity was further assessed by examining the Mermin inequality for multiphoton GHZ states, for which significant violations were observed for modeled states of up to five photons. This model can be useful for simulating mixed entangled multiphoton systems, such as are used in realistic quantum information processing devices, or for explorations of the quantum-classical boundary.

%%%%%%%%%%%%%%%%%%%%%%%%%%%%%%%%%%%%%%%%%%%%%%%%

\printcredits

\section*{Data availability}

No data was used for the research described in this article.

\section*{Funding sources}
This work was funded by the Office of Naval Research under Grants {N00014-18-1-2107} and {N00014-23-1-2115}.

%%%%%%%%%%%%%%%%%%%%%%%%%%%%%%%%%%%%%%%%%%%%%%%%%

\appendix

\section{Proof of Equation (\ref{eqn:Bogoliubov})}
\label{app:Agnes}

We show that, for Hilbert space operators $\hat{X}$, $\hat{Y}$, and $\hat{A}$, if $\hat{A}$ is skew Hermitian then
\begin{equation}
\hat{Y} = e^{-\hat{A}} \hat{X} e^{\hat{A}} = \sum_{k=0}^{\infty} \frac{\hat{C}^{(k)}}{k!} \; ,
\label{eqn:Angus}
\end{equation}
where $\hat{C}^{(0)} = \hat{X}$ and, for $k \in \mathbb{N}$, $\hat{C}^{(k)} = [\hat{C}^{(k-1)}, \, \hat{A}]$.  We begin by writing
\begin{equation}
e^{-\hat{A}} \hat{X} e^{\hat{A}} = \sum_{m=0}^{\infty} \frac{(-\hat{A})^m}{m!} \sum_{n=0}^{\infty} \frac{1}{n!} \, \hat{X}\hat{A}^n
\label{eqn:Alice}
\end{equation}
and proving by induction that
\begin{equation}
\hat{X} \hat{A}^n = \sum_{k=0}^{n} \binom{n}{k} \hat{A}^{n-k} \, \hat{C}^{(k)} \; .
\label{eqn:Fred}
\end{equation}

Clearly Eqn.\ (\ref{eqn:Fred}) holds for $n=0$.  For the induction step, let $n > 0$ and note that
\begin{equation}
\begin{split}
\hat{a}_i \hat{A}^n &= \hat{a}_i \hat{A}^{n-1} \hat{A} \\
&= \sum_{k=0}^{n-1} \binom{n-1}{k} \hat{A}^{n-1-k} \, \hat{C}^{(k)} \hat{A} \\
&= \sum_{k=0}^{n-1} \binom{n-1}{k} \hat{A}^{n-1-k} \, \left( \hat{A} \hat{C}^{(k)} + [\hat{C}^{(k)}, \hat{A}] \right) \\
&= \sum_{k=0}^{n-1} \binom{n-1}{k} \hat{A}^{n-1-k} \, \left( \hat{A} \hat{C}^{(k)} + \hat{C}^{(k+1)} \right) \\
&= \sum_{k=0}^{n-1} \binom{n-1}{k} \hat{A}^{n-k} \hat{C}^{(k)} \\
&\quad + \sum_{k=0}^{n-1} \binom{n-1}{k} \hat{A}^{n-(k+1)} \hat{C}^{(k+1)} \; .
\end{split}
\end{equation}
By reindexing, the last term may be written
\begin{multline}
\sum_{k=0}^{n-1} \binom{n-1}{k} \hat{A}^{n-(k+1)} \hat{C}^{(k+1)} \\
= \sum_{k=1}^{n-1} \binom{n-1}{k-1} \hat{A}^{n-k} \hat{C}^{(k)} + \binom{n}{n} \hat{A}^0 \hat{C}^{(n)} \; .
\end{multline}
Finally, using Pascal's rule for $1 \le k \le n-1$,
\begin{equation}
\binom{n-1}{k} + \binom{n-1}{k-1} = \binom{n}{k} \; ,
\end{equation}
we find
\begin{equation}
\begin{split}
\hat{X} \hat{A}^n &= \sum_{k=0}^{n-1} \binom{n}{k} \hat{A}^{n-k} \hat{C}^{(k)} + \binom{n}{n} \hat{A}^0 \hat{C}^{(n)} \\
&= \sum_{k=0}^{n} \binom{n}{k} \hat{A}^{n-k} \hat{C}^{(k)} \; .
\end{split}
\end{equation}
This proves Eqn.\ (\ref{eqn:Fred}).  Substituting Eqn.\ (\ref{eqn:Fred}) into Eqn. (\ref{eqn:Alice}) now gives
\begin{equation}
\begin{split}
\hat{Y} &= \sum_{m=0}^{\infty} \frac{(-\hat{A})^m}{m!} \sum_{n=0}^{\infty} \frac{1}{n!} \sum_{k=0}^{n} \binom{n}{k} \hat{A}^{n-k} \hat{C}^{(k)} \\
&= \sum_{m=0}^\infty \sum_{n=0}^\infty \sum_{k=0}^{n} \frac{(-\hat{A})^m \hat{A}^{n-k}}{m! \, n!} \binom{n}{k}\hat{C}^{(k)}\\
&= \sum_{m=0}^\infty \sum_{j=0}^\infty \sum_{k=0}^{\infty} \frac{(-\hat{A})^m \hat{A}^{j}}{m! \, (j+k)!} \binom{j+k}{k}\hat{C}^{(k)}\\
&= \sum_{p=0}^\infty \sum_{j=0}^p \sum_{k=0}^{\infty} \frac{(-\hat{A})^{p-j} \hat{A}^{j}}{(p-j)! \, (j+k)!} \binom{j+k}{k}\hat{C}^{(k)}\\
&= \sum_{p=0}^\infty \sum_{k=0}^{\infty} \frac{(\hat{A}-\hat{A})^p}{p! \, k!} \hat{C}^{(k)} = \sum_{k=0}^{\infty} \frac{\hat{C}^{(k)}}{k!} \; .
\end{split}
\end{equation}
This proves Eqn.\ (\ref{eqn:Angus}).

%%%%%%%%%%%%%%%%%%%%%%%%%%%%%%%%%%%%%%%%%%%%%%%%%

\section{Proof of Equation (\ref{eqn:Damien})}
\label{app:Angus}

The first-order term in $\hat{b}_i$ is given by
\begin{equation}
\begin{split}
\hat{C}_i^{(1)} &= \frac{1}{2} \sum_{\ell,j} \xi_{\ell j} [\hat{a}_i, \; \hat{a}_{\ell}^\dagger \hat{a}_j^\dagger] \\
&= \frac{1}{2} \sum_{\ell,j} \xi_{\ell j} \left( [\hat{a}_i, \hat{a}_{\ell}^\dagger] \hat{a}_j^\dagger + \hat{a}_{\ell}^\dagger [\hat{a}_i, \hat{a}_j^\dagger] \right) \\
&= \frac{1}{2} \sum_{j} \xi_{ij} \hat{a}_j^\dagger + \frac{1}{2} \sum_{\ell} \xi_{\ell i} \hat{a}_{\ell}^\dagger \\
&= \frac{1}{2} \sum_{j} \xi_{ij} \hat{a}_j^\dagger + \frac{1}{2} \sum_{\ell} \xi_{i \ell} \hat{a}_{\ell}^\dagger \\
&= \sum_{j} \xi_{ij} \hat{a}_j^\dagger \; .
\end{split}
\end{equation}
Using this result, $\hat{C}_i^{(2)}$ is found to be
\begin{equation}
\begin{split}
\hat{C}_i^{(2)} &= \sum_{j} \xi_{ij} [\hat{a}_j^\dagger, \hat{A}] \\
&= \sum_{j} \xi_{ij} [\hat{a}_j, \hat{A}]^\dagger \\
&= \sum_{j} \xi_{ij} \sum_{k} \xi_{jk}^* \hat{a}_k \; ,
\end{split}
\end{equation}
and $\hat{C}_i^{(3)}$ is found similarly to be
\begin{equation}
\hat{C}_i^{(3)} = \sum_{jk\ell} \xi_{ij} \xi_{jk}^* \xi_{k\ell} \hat{a}_{\ell}^\dagger \; .
\end{equation}

The general result may be written in matrix form more compactly.  Writing $\vec{C}^{(k)} = [C_1^{(k)}, \ldots, C_{2d}^{(k)}]^\mathsf{T}$, it can be shown that
\begin{equation}
\hat{\vec{C}}^{(k)} = 
\begin{cases}
\mtx{R}^k \,\hat{\vec{a}} & \quad \text{if $k$ is even;} \\
\mtx{R}^k \mtx{Q} \,\hat{\vec{a}}^\dagger & \quad \text{if $k$ is odd.} \\ 
\end{cases}
\label{eqn:Charlene}
\end{equation}
This is easily proven by induction following the examples above.  Now, using Eqns.\ (\ref{eqn:Bogoliubov}) and (\ref{eqn:Charlene}), we recover the familiar result
\begin{equation}
\begin{split}
\hat{\vec{b}} &= \sum_{\ell=0}^{\infty} \frac{\hat{\vec{C}}^{(2\ell)}}{(2\ell)!} + \sum_{\ell=0}^{\infty} \frac{\hat{\vec{C}}^{(2\ell+1)}}{(2\ell+1)!} \\
&= \sum_{\ell=0}^{\infty} \frac{\mtx{R}^{2\ell} \hat{\vec{a}}}{(2\ell)!} + \sum_{\ell=0}^{\infty} \frac{\mtx{R}^{2\ell+1} \mtx{Q} \hat{\vec{a}}^\dagger}{(2\ell+1)!} \\
&= (\cosh\mtx{R}) \hat{\vec{a}} + (\sinh\mtx{R}) \mtx{Q} \hat{\vec{a}}^\dagger \; .
\end{split}
\end{equation}
This proves Eqn.\ (\ref{eqn:Damien}).

%%%%%%%%%%%%%%%%%%%%%%%%%%%%%%%%%%%%%%%%%%%%%%%%%

\section{Second-order Approximation for $n \ge 3$}
\label{app:Beatrice}

Let $\hat{b}_i^{(p)}$ denote the order-$p$ approximation of $\hat{b}_i$, defined as
\begin{equation}
\hat{b}_i^{(p)} = \sum_{k=0}^{p} \frac{\hat{C}_i^{(k)}}{k!} \; .
\label{eqn:bp}
\end{equation}
Although $\hat{b}_i^{(p)}$ no longer obeys the bosonic commutation relations, it may nevertheless serve as a useful tool for the purpose of approximating expectation values.

The first-order term $\hat{C}_i^{(1)}$ is simple enough to be calculated explicitly.  Since $\hat{C}_i^{(0)} = \hat{a}_i$, we note that
\begin{equation}
\begin{split}
\hat{C}_i^{(1)} &= \frac{1}{n!} \sum_{i_1,\ldots,i_n} \xi_{i_1,\ldots,i_n}  [\hat{a}_i, \; \hat{a}_{i_1}^\dagger \cdots \hat{a}_{i_n}^\dagger] \; .
\end{split}
\end{equation}
The commutator contains $n$ terms, given by
\begin{equation}
[\hat{a}_i, \; \hat{a}_{i_1}^\dagger \cdots \hat{a}_{i_n}^\dagger] = 
\sum_{m=1}^{n} \prod_{\ell=1}^{m-1} \hat{a}_{i_\ell}^\dagger \; [\hat{a}_{i}, \hat{a}_{i_m}^\dagger] \prod_{\ell'=m+1}^{n} \hat{a}_{i_{\ell'}}^\dagger \; .
\end{equation}
Since $[\hat{a}_{i}, \hat{a}_{i_m}^\dagger] = \delta_{i,i_m} \hat{1}$ and $\mtx{\xi}$ is symmetric, we find that
\begin{equation}
\hat{C}_i^{(1)} = \frac{1}{(n-1)!} \sum_{i_2,\ldots,i_n} \xi_{i,i_2,\ldots,i_n}  \hat{a}_{i_2}^\dagger \cdots \hat{a}_{i_n}^\dagger \; .
\end{equation}

The second-order term $\hat{C}_i^{(2)}$ is somewhat more complicated to compute.  We begin with
\begin{equation}
\hat{C}_i^{(2)} = \frac{1}{(n-1)!} \sum_{i_2,\ldots,i_n} \xi_{i,i_2,\ldots,i_n} [\hat{a}_{i_2}^\dagger \cdots \hat{a}_{i_n}^\dagger, \; \hat{A}]
\end{equation}
and note that
\begin{equation}
[\hat{a}_{i_2}^\dagger \cdots \hat{a}_{i_n}^\dagger, \hat{A}] 
= \frac{1}{n!} \sum_{j_1,\ldots,j_n} \xi_{j_1,\ldots,j_n}^* [\hat{a}_{j_1} \cdots \hat{a}_{j_n}, \; \hat{a}_{i_2}^\dagger \cdots \hat{a}_{i_n}^\dagger] \; .
\end{equation}
The commutator contains $n-1$ terms, given by
\begin{multline}
[\hat{a}_{j_1} \cdots \hat{a}_{j_n}, \; \hat{a}_{i_2}^\dagger \cdots \hat{a}_{i_n}^\dagger] \\
= \sum_{m=2}^{n} \prod_{\ell=2}^{m-1} \hat{a}_{i_\ell}^\dagger \left[ \prod_{k=1}^{n} \hat{a}_{j_k}, \; \hat{a}_{i_m}^\dagger \right] \prod_{\ell'=m+1}^{n} \hat{a}_{i_{\ell'}}^\dagger \; ,
\end{multline}
and the inner commutator may be further decomposed into $n$ terms such that
\begin{equation}
\left[ \prod_{k=1}^{n} \hat{a}_{j_k}, \; \hat{a}_{i_m}^\dagger \right] = \sum_{p=1}^{n} \delta_{j_p,i_m} \prod_{s \neq p} \hat{a}_{j_s} \; .
\end{equation}

Bringing in the sum over $j_1, \ldots, j_n$ and using the symmetry of $\mtx{\xi}$ yields
\begin{equation}
\begin{split}
\sum_{j_1,\ldots,j_n} \xi_{j_1,\ldots,j_n}^* &\left[ \prod_{k=1}^{n} \hat{a}_{j_k}, \; \hat{a}_{i_m}^\dagger \right] \\
&= \sum_{j_1,\ldots,j_n} \xi_{j_1,\ldots,j_n}^* \sum_{p=1}^{n} \delta_{j_p,i_m} \prod_{s \neq p} \hat{a}_{j_s} \\
&= n \sum_{j_2,\ldots,j_n} \xi_{i_m,j_2,\ldots,j_n}^* \prod_{s=2}^{n} \hat{a}_{j_s} \; ,
\end{split}
\end{equation}
so that
\begin{equation}
[\hat{a}_{i_2}^\dagger \cdots \hat{a}_{i_n}^\dagger, \hat{A}] = \sum_{j_2,\ldots,j_n} \sum_{m=2}^{n} \frac{\xi_{i_m,j_2,\ldots,j_n}}{(n-1)!} \; \hat{B}_{j_2,\ldots,j_n}^{i_2,\ldots,i_n} \; ,
\end{equation}
where
\begin{equation}
\hat{B}_{j_2,\ldots,j_n}^{i_2,\ldots,i_n} = \prod_{\ell=2}^{m-1} \hat{a}_{i_\ell}^\dagger \prod_{s=2}^{n} \hat{a}_{j_s} \prod_{\ell'=m+1}^{n} \hat{a}_{i_{\ell'}}^\dagger \; .
\end{equation}
Since the indices $i_2,\ldots,i_3$ can be relabeled within the sum over $m$, we  conclude
\begin{equation}
\hat{C}_i^{(2)} = \sum_{i_2,\ldots,i_n} \frac{\xi_{i,i_2,\ldots,i_n}}{(n-1)!} \sum_{j_2,\ldots,j_n} \frac{\xi_{i_2,j_2,\ldots,j_n}^*}{(n-1)!} \hat{D}^{i_3,\ldots,i_n}_{j_2,\ldots,j_n} \; ,
\end{equation}
where
\begin{equation}
\hat{D}^{i_3,\ldots,i_n}_{j_2,\ldots,j_n} = \sum_{m=2}^{n} \prod_{\ell=3}^{m} \hat{a}_{i_\ell}^\dagger \prod_{s=2}^{n} \hat{a}_{j_s} \prod_{\ell'=m+1}^{n} \hat{a}_{i_{\ell'}}^\dagger \; .
\end{equation}

%%%%%%%%%%%%%%%%%%%%%%%%%%%%%%%%%%%%%%%%%%%%%%%%%

\section{Approximating $\ket{\Psi^{(p)}}$ with $\vec{b}^{(p)}$}
\label{app:Charlene}

The pinched annihilation operator, $\hat{b}_i$, may be approximated to order $p$ as
\begin{equation}
\begin{split}
(\hat{S}^{(p)})^\dagger \hat{a}_i \hat{S}^{(p)} &= \sum_{k'=0}^{p} \frac{(-\hat{A})^{k'}}{(k')!} \hat{a}_i \sum_{\ell=0}^{p} \frac{\hat{A}^\ell}{\ell!} \\
&= \sum_{\ell=0}^{p} \sum_{k=\ell}^{\ell+p} \frac{(-\hat{A})^{k-\ell} \hat{a}_i \hat{A}^\ell}{(k-\ell)! \, \ell!} \\
&= \sum_{k=0}^{p} \frac{1}{k!} \sum_{\ell=0}^{k} \binom{k}{\ell} (-\hat{A})^{k-\ell} \hat{a}_i \hat{A}^{\ell} \\
&\quad + \sum_{k=p+1}^{2p} \sum_{\ell=k-p}^{p} \frac{(-\hat{A})^{k-\ell} \hat{a}_i \hat{A}^{\ell}}{(k-\ell)! \, \ell!} \\
&= \hat{b}_i^{(p)} + \hat{\varepsilon}_i^{(p)} \; , 
\end{split}
\end{equation}
where we have used the expression for $\hat{C}^{(k)}_i$ in Eqn.\ (\ref{eqn:Ck}) and that of $\hat{b}^{(p)}_i$  from Eqn.\ (\ref{eqn:bp}) for the first term.  The remaining second term is defined to be
\begin{equation}
\hat{\varepsilon}_i^{(p)} = \sum_{k=p+1}^{2p} \sum_{\ell=k-p}^{p} \frac{(-\hat{A})^{k-\ell} \hat{a}_i \hat{A}^{\ell}}{(k-\ell)! \, \ell!} \; .
\end{equation}

If we suppose $\mtx{\xi}$ is bounded (i.e., $|\xi_{i_1,\ldots,i_n}| < r$ for all $i_1,\ldots,i_n$ and some $r > 0$) and $r \ll 1$ is small, then the expectation value of $\hat{\varepsilon}^{(p)}$ with respect to some quantum state will be small relative to $\hat{b}^{(p)}_i$ since
\begin{equation}
\begin{split}
|\braket{\hat{\varepsilon}^{(p)}_i}| &\le \sum_{k=p+1}^{2p} \sum_{\ell=k-p}^{p} \frac{\left| \braket{ (-\hat{A})^{k-\ell} \hat{a}_i \hat{A}^{\ell} }\right|}{(k-\ell)! \, \ell!} \\
&= \sum_{k=p+1}^{2p} \mathcal{O}(r^k) = \mathcal{O}\Bigl( \min\{r^{p+1},r^{2p}\} \Bigr) \; .
\end{split}
\end{equation}

%%%%%%%%%%%%%%%%%%%%%%%%%%%%%%%%%%%%%%%%%%%%%%%%%

% \bibliographystyle{cas-model2-names}
% \bibliography{refs}

\end{document}